\documentstyle[12pt,psfig]{article}

\begin{document}
\begin{titlepage}

\title{Exactly solvable multichannel Kondo-lattice model}

\vspace{3cm}
	
\author{C.-A. Piguet, D.F. Wang and C. Gruber} 
	
\date{March 1996}
	
\maketitle
 
\begin{center}
Institut de Physique Th\'eorique \\
Ecole Polytechnique F\'ed\'erale de Lausanne \\
PHB-Ecublens, CH-1015 Lausanne, Switzerland
\end{center}

\vspace{3cm}

In this work, a multichannel Kondo-lattice model is studied
in the thermodynamic limit. The conduction band is described 
by a constant hopping amplitude between any pair of lattice sites.
For this system we have obtained the exact thermodynamical properties and the
ground-state energies. In the limit of strong
interaction between the conduction electrons and the impurity
spins, the wavefunctions take the Jastrow product form. 
\end{titlepage}
\newpage

\section{Introduction}

The Kondo-lattice model consists of one or several 
electronic conduction bands (channels)
interacting with some impurities through a spin exchange term \cite{lee,schl}.
This model is relevant for the studies of the so-called Kondo-insulators and
heavy fermion rare-earth compounds. There are considerable recent interests in
the study of magnetic impurity effects 
in conduction electrons [3-7].  

For one dimensional Kondo-lattice, various numerical and theoretical investigations
were carried out [8-13]. 
It was found that at half-filling any arbitrarily small interaction
would make the system an insulator [8-13]. One interesting open question
is whether dimensionality plays a role in this Kondo-insulator, similar to
the Mott-insulator due to electron-electron correlation. 

In this work, we consider a Kondo-lattice model with multichannel conduction
bands. In the extreme situation, where the conduction electrons 
have constant hopping amplitude between any pair of lattice sites, 
we show that one can rigorously construct the free energy of the system
in the thermodynamic limit. This generalizes the previous result \cite{gruber}
to the multichannel case. In particular, at zero temperature, 
our exact solutions demonstrate interesting metal-insulator phase 
transitions in the system when the filling numbers of the electrons
vary. In the limit of infinite interaction, the ground-states take 
Jastrow product form. It remains unclear how to define effective masses
of electrons in this simplified model, and how to study heavy mass
for the electrons induced by the impurity spins. 
Nevertheless, in this simple case, one can write the
exact solutions for the ground-state and the free energy, and we 
report these results and their derivations in the following
sections. 
 
\section{The multichannel Kondo-lattice model}

The hamiltonian of the $M$-channel Kondo model on a lattice with exactly one
impurity spin at each site and an arbitrary number of orbital conduction electrons
has the form
\begin{equation}
H=H_{0}+JV
\label{hamiltonian}
\end{equation}
where $H_{0}$ is the kinetic energy of the electrons, $J$ is the coupling
constant, and the spin-spin interaction term $V$ is given by
\begin{equation}
V=\sum_{i=1}^{L}
\sum_{m=1}^{M}\vec {S}_{i}^{f}\cdot\vec{S}_{im}^{e}
\end{equation}
with $\vec {S}_{i}^{f}$ the spin operator for the impurity at the site $i$, $L$
the number of sites, $m$ labels the channels, and $\vec{S}_{im}^{e}$ the spin
operator for the electron at the site $i$ in the $m$-th band.

Introducing the creation and annihilation fermionic operators $(c_{i\sigma
m}^{+},c_{i\sigma m})$ for an electron at site $i$, in the $m$-th band, with
spin $\sigma$, we have 
\begin{equation}
\vec{S}_{im}^{e}=\frac{1}{2}\sum_{\alpha,\beta=\uparrow ,\downarrow}
c_{i\alpha m}^{+}\vec{\sigma}_{\alpha\beta}
c_{i\beta m}
\end{equation}
where $\vec{\sigma}$ are the Pauli matrices. Similarly we introduce creation 
and annihilation fermionic operators $(f_{i\sigma}^{+},f_{i\sigma})$ 
for an impurity at site $i$ with spin $\sigma$ and
express the spin of the impurity at site $i$ as
\begin{equation}
\vec{S}_{i}^{f}=\frac{1}{2}\sum_{\alpha,\beta=\uparrow , \downarrow}
f_{i\alpha}^{+}\vec{\sigma}_{\alpha\beta}
f_{i\beta}.
\label{spin}
\end{equation}
Clearly the $c$-operators commute with the $f$-operators.

Let ${\cal H}$ denote the Hilbert space with an arbitrary (0, 1, or 2) number of
impurities at each site. The condition that there is exactly one
impurity at each site means that we restrict ourselves to the subspace $\bar{{\cal H}}$
where
\begin{equation}
\sum_{\sigma=\uparrow,\downarrow}f_{i\sigma}^{+}f_{i\sigma}=1,\;\;{\rm for\;
all}\;\;i=1,\ldots,L.
\end{equation}
 
In our model we consider a kinetic energy $H_{0}$ with constant hopping
amplitude $t$, i.e.
\begin{equation}
H_{0}=-t\sum_{1\leq i\neq j\leq L}\sum_{\sigma=\uparrow,\downarrow}\sum_{m=1}^{M}
c_{i\sigma m}^{+}c_{j\sigma m}.
\end{equation}
Because of the special form of $H_{0}$, the dimensionality of the lattice is
irrelevant and the system is basically one dimensional. In the next section we
shall thus discuss the one-dimensional system in the thermodynamic limit $L\rightarrow\infty$. 

Let us introduce the Fourier transform of the electronic operators by $c_{k\sigma
m}^{+}=\frac{1}{\sqrt{L}}\sum_{j=1}^{L}e^{ikj}c_{j\sigma m}^{+}$ and $c_{k\sigma
m}=\frac{1}{\sqrt{L}}\sum_{j=1}^{L}e^{-ikj}c_{j\sigma m}$ where $k$ is in the
first Brillouin  zone. In terms of these
operators, the kinetic part $H_{0}$ of the hamiltonian is diagonal
\begin{equation}
H_{0}=\sum_{k\in
FBZ}\sum_{\sigma=\uparrow,\downarrow}\sum_{m=1}^{M}\epsilon(k)c_{k\sigma
m}^{+}c_{k\sigma m}
\end{equation}
with the dispersion relation
\begin{equation}
\epsilon(k)=-t'L\delta_{0k}+t
\end{equation}
and $t'=t$. However, in the following, we will consider $t'$ and $t$ as two
independent parameters.

\section{Cluster expansion}

Although we are not able to diagonalize the hamiltonian
(\ref{hamiltonian}) for the finite system, we can, as in
\cite{gruber} or \cite{dongen}, use a cluster expansion 
to obtain the grand-canonical potential $\Omega/L$
in the thermodynamic limit. The potential $\Omega$ is defined by
\begin{equation}
\Omega=-\frac{1}{\beta}\ln {\cal Z}
\end{equation}
with ${\cal Z}$ the grand-canonical partition function 
\begin{equation}
{\cal Z}=Tr_{\bar{\cal H}} e^{-\beta K}
\end{equation}
where
\begin{eqnarray}
K&=&H-\sum_{m=1}^{M}\mu_{m}\hat{N}_{m}^{e}=K_{0}+JV\nonumber\\
K_{0}&=&H_{0}-\sum_{m=1}^{M}\mu_{m}\hat{N}_{m}^{e}.
\end{eqnarray}
$\mu_{m}$ and $\hat{N}_{m}^{e}$ are the chemical potential and the number of
electrons operator of the $m$-th band.

In the definition of ${\cal Z}$, the trace is performed over the Hilbert space
$\bar{\cal H}$ where at each site there is exactly one impurity. It is however more
convenient to work in the Hilbert space ${\cal H}$.
For this, following \cite{coleman}, we introduce the impurity
isospin at site $i$ by
\begin{equation}
\vec{\tau}_{i}^{f}=\frac{1}{2}\sum_{\alpha,\beta=\uparrow,\downarrow}
\tilde{f}_{i\alpha}^{+}\vec{\sigma}_{\alpha\beta}\tilde{f}_{i\beta}
\label{isospin}
\end{equation}
where $\tilde{f}_{i}$ is the Nambu spinor:
\begin{equation}
\tilde{f}_{i}=\left(\begin{array}{c}
f_{i\uparrow}\\
f^{+}_{i\downarrow} \end{array} \right).
\end{equation}

Let us consider the sum of the spin and of the isospin at site $i$:
$\vec{S}_{i}^{f}+\vec{\tau}_{i}^{f}$. In the subspace where at site $i$ we have
exactly one impurity, $\vec{S}_{i}^{f}$ acts as a $1/2$ spin operator and
$\vec{\tau}_{i}^{f}$ as the zero operator. In the subspace where at site $i$ we
have zero or two impurities, $\vec{S}_{i}^{f}$ acts as the zero operator and
$\vec{\tau}_{i}^{f}$ as a $1/2$ spin operator (the zero impurity spin state
corresponding to a down spin and the two impurities state corresponding to a up
spin). Thus, if we replace in $K$ the spin by the sum of the spin and of the
isospin, we have $2^{L}$ different subspaces in which this new operator acts
similarly. This allows us to rewrite the partition function with a trace on the
whole Hilbert space:
\begin{equation}
{\cal Z}=Tr_{\bar{\cal H}}e^{-\beta K(\vec{S}^{f})}=\frac{1}{2^{L}} 
Tr_{\cal H}e^{-\beta K(\vec{S}^{f}+\vec{\tau}^{f})}.
\end{equation}

Instead of using the expressions (\ref{spin}) and (\ref{isospin}) for the
definitions of the spin and of the isospin, we prefer to use Majorana fermions.
For this, we decompose the $f$-spinor in real and imaginary part by:
\begin{equation}
\left(\begin{array}{c}
f_{i\uparrow}\\
f_{i\downarrow}\end{array}\right)=\frac{1}{\sqrt{2}}
\left(\begin{array}{c}
-\eta_{i}^{1}+i\eta_{i}^{2}\\
\eta_{i}^{3}+i\eta_{i}^{0}\end{array}\right)
\end{equation}
where $\eta_{i}^{0}$, $\eta_{i}^{1}$, $\eta_{i}^{2}$ and $\eta_{i}^{3}$ are the
Majorana fermions.
These operators are hermitian and satisfy anticommutation relations: $\{\eta_{i}^{a},
\eta_{j}^{b}\}=\delta_{ij}\delta_{ab}$. We define their Fourier transform
\begin{equation}
\eta_{k}^{a}=\frac{1}{\sqrt{L}}\sum_{j=1}^{L}e^{ikj}\eta_{j}^{a}
\end{equation}
which satisfy the anticommutation relations
$\{\eta_{k}^{a},\eta_{k'}^{b}\}=\delta_{k,-k'}\delta_{ab}$. In terms of
the Majorana fermions the sum of the spin and of the isospin has a simple form:
\begin{equation}
\vec{S}_{i}^{f}+\vec{\tau}_{i}^{f}=-\frac{i}{2}\vec{\eta}_{i}\wedge\vec{\eta}_{i}
\end{equation}
where $\vec{\eta}_{i}=(\eta_{i}^{1},\eta_{i}^{2},\eta_{i}^{3})$. Using the
commutation relations of the Pauli matrices
($\vec{\sigma}=-(i/2)\vec{\sigma}\wedge\vec{\sigma}$), we can rewrite the spin
exchange term as
\begin{equation}
V=-\frac{1}{8}\sum_{i=1}^{L}\sum_{\alpha,\beta=\uparrow,\downarrow}\sum_{m=1}^{M}
c_{i\alpha m}^{+}c_{i\beta m}(\vec{\sigma}\wedge\vec{\sigma})_{\alpha\beta}
(\vec{\eta}_{i}\wedge\vec{\eta}_{i}).
\end{equation} 

Let us now apply the cluster expansion method to the thermodynamic potential:
\begin{equation}
\Omega=\Omega_{0}-\frac{1}{\beta}\sum_{n=1}^{\infty}J^{n}W_{n}
\end{equation}
with $\Omega_{0}=-\frac{1}{\beta}\ln\left(\frac{1}{2^{L}}Tr_{\cal H}e^{-\beta
K_{0}}\right)$ and
\begin{equation}
W_{n}=(-1)^{n}\int_{0}^{\beta}d\tau_{1}\int_{0}^{\tau_{1}}d\tau_{2}\cdots
\int_{0}^{\tau_{n-1}}d\tau_{n}<V(\tau_{1})V(\tau_{2})\cdots
V(\tau_{n})>^{c}_{0}.
\label{expansion}
\end{equation}
$V(\tau)$ is the free Heisenberg representation of the potential, 
and $<\cdots>^{c}_{0}$ denotes the grand-canonical average of the system without
interaction, taken only over connected diagrams:
\begin{eqnarray}
V(\tau)&=&e^{\tau K_{0}}Ve^{-\tau K_{0}}\\
<V(\tau_{1})V(\tau_{2})\cdots V(\tau_{n})>_{0}&=&Tr_{\cal H}(\rho_{0}
V(\tau_{1})V(\tau_{2})\cdots V(\tau_{n}))
\end{eqnarray}
where $\rho_{0}=e^{-\beta K_{0}}/Tr_{\cal H} e^{-\beta K_{0}}$ is the density
matrix of the model without interaction.

Since $K_{0}$ is diagonal in the Fourier space, 
we have for the electronic operators
\begin{eqnarray}
c_{k\sigma m}^{+}(\tau)&=&e^{(\epsilon(k)-\mu_{m})\tau}c_{k\sigma m}^{+}\\  
c_{k\sigma m}(\tau)&=&e^{-(\epsilon(k)-\mu_{m})\tau}c_{k\sigma m}.  
\end{eqnarray}
Let us remark that $c_{k\sigma m}^{+}(\tau)$ and $c_{k\sigma m}(\tau)$ are
no more adjoint to each other. Moreover the Majorana fermions commute with
$K_{0}$ and thus they do not depend on $\tau$:
\begin{equation}
\vec{\eta}_{i}(\tau)=\vec{\eta}_{i}.
\end{equation}
This allows us to write explicitly the interaction term $V(\tau)$ in the
Fourier space:
\begin{equation}
V(\tau)=-\frac{1}{8L}\sum_{k_{1}k_{2}k_{3}k_{4}\in
FBZ}\sum_{\alpha,\beta=\uparrow,\downarrow}
\sum_{m=1}^{M}\delta_{k_{1}-k_{2}+k_{3}+k_{4},0}c_{k_{1}\alpha m}^{+}(\tau)c_{k_{2}\beta m}(\tau)
(\vec{\sigma}\wedge\vec{\sigma})_{\alpha\beta}(\vec{\eta}_{k_{3}}\wedge\vec{\eta}_{k_{4}}).
\end{equation}

In order to calculate the grand-canonical averages, we use Wick's theorem, which 
tells us that any average of product of operators can be
expressed as a sum of product of averages of two operators. We can compute these
averages explicitly:
\begin{eqnarray}
<c_{k\sigma m}^{+}\eta_{k}^{a}>_{0}&=&
<c_{k\sigma m}\eta_{k}^{a}>_{0}=0\nonumber\\
<c_{k\sigma m}^{+}c_{k'\sigma' m'}^{+}>_{0}&=& 
<c_{k\sigma m}c_{k'\sigma' m'}>_{0}=0\nonumber\\
<c_{k\sigma m}^{+}c_{k'\sigma'
m'}>_{0}&=&\delta_{kk'}\delta_{\sigma\sigma'}\delta_{mm'}
\frac{e^{-\beta(\epsilon(k)-\mu_{m})}}{1+e^{-\beta(\epsilon(k)-\mu_{m})}}\nonumber\\
<c_{k\sigma m}c_{k'\sigma'
m'}^{+}>_{0}&=&\delta_{kk'}\delta_{\sigma\sigma'}\delta_{mm'}
\frac{1}{1+e^{-\beta(\epsilon(k)-\mu_{m})}}\nonumber\\
<\eta_{k}^{a}\eta_{k'}^{b}>_{0}&=&\frac{1}{2}\delta_{k,-k'}\delta_{ab}.
\end{eqnarray}
Since $\epsilon(k)=-t'L\delta_{0k}+t$, these averages do not depend
on $t'$ except those with momentum zero. However, in the thermodynamic limit,
and for $t'>0$, we have
\begin{eqnarray}
\lim_{L \rightarrow \infty} <c_{0\sigma l}^{+}(\tau_{i})
c_{0\sigma l}(\tau_{j})>_{0}&=&\left\{ \begin {array}{ll}
1\;\;\; & \tau_{i}=\tau_{j} \\
0\;\;\; & \tau_{i}>\tau_{j} 
\end {array} \right.\nonumber\\ 
\lim_{L \rightarrow \infty} <c_{0\sigma l}(\tau_{i})
c_{0\sigma l}^{+}(\tau_{j})>_{0}&=&\left\{ \begin {array}{ll}
0\;\;\; & \tau_{i}-\tau_{j}<\beta \\
1\;\;\; & \tau_{i}-\tau_{j}=\beta.
\end {array} \right.
\label{pos} 
\end{eqnarray}
On the other hand, for $t'<0$,
\begin{eqnarray}
\lim_{L \rightarrow \infty} <c_{0\sigma l}^{+}(\tau_{i})
c_{0\sigma l}(\tau_{j})>_{0}&=&
\left\{ \begin {array}{ll}
1\;\;\; & \tau_{i}-\tau_{j}=\beta \\
0\;\;\; & \tau_{i}-\tau_{j}<\beta 
\end {array} \right. \nonumber\\
\lim_{L \rightarrow \infty} <c_{0\sigma l}(\tau_{i})
c_{0\sigma l}^{+}(\tau_{j})>_{0}&=&
\left\{ \begin {array}{ll}
1\;\;\; & \tau_{i}=\tau_{j} \\
0\;\;\; & \tau_{i}>\tau_{j}. 
\end {array} \right.
\label{neg} 
\end{eqnarray}

Therefore from Eq.(\ref{expansion}), (\ref{pos}), (\ref{neg}), the
coefficients $W_{n}$, $n\geq 1$, become independent of $t'$ in the thermodynamic
limit. Assuming we can permute the limit $L\rightarrow\infty$ and the sum over
$n$, we are left with the corresponding expression for the model, with $t'=0$:
\begin{eqnarray}
\Omega&=&-2MLt'\theta(t')-\frac{L}{\beta}\ln\left(2\prod_{m=1}^{M}(1+
e^{\beta(\mu_{m}-t)})^{2}\right)-\frac{1}{\beta}\sum_{n=1}^{\infty}J^{n}W_{n}(t'=0)+O(1)\nonumber\\
&=&-2MLt'\theta(t')+\Omega_{0}(t'=0)
-\frac{1}{\beta}\sum_{n=1}^{\infty}J^{n}W_{n}(t'=0)+O(1)\nonumber\\
&=&-2MLt'\theta(t')+\Omega(t'=0)+O(1)
\label{omega}
\end{eqnarray}
where $\theta(t')=1$ if $t'>0$ and $\theta(t')=0$ if $t'<0$. $\Omega(t'=0)$
is the thermodynamic potential for the system with $t'=0$.

We have thus shown that in the thermodynamic limit the potential is made of two
contributions in which the dependance in $t'$ and $J$ separate. The first
contribution is trivial and we can compute explicitly the second one, since it
is site by site diagonal. In the next section, this fact will be used to
obtain the ground-state energy.

\section{Ground-state energy and phase transitions}

In this section, we study the zero temperature properties of 
the system. The ground-state energies will be obtained as
functions of electron filling numbers. As the filling numbers 
vary, one finds metal-insulator phase transitions. 

From Eq.(\ref{omega}), the number of electrons in each band and the energy are
given by
\begin{equation}
N_{m}=-\left.\frac{\partial\Omega}{\partial \mu_{m}}\right)_{\beta}=
-\left.\frac{\partial\Omega(t'=0)}{\partial
\mu_{m}}\right)_{\beta,\mu_{1},\ldots,\mu_{m-1},\mu_{m+1},\ldots,\mu_{M}}=N_{m}(t'=0)
\end{equation}
\begin{equation}
E=\sum_{m=1}^{M}\mu_{m}N_{m}+\left.\frac{\partial\beta\Omega}{\partial\beta}\right)
_{\mu_{1},\ldots,\mu_{M}}=-2t'ML\theta(t')+t\sum_{m=1}^{M}N_{m}+E^{int},
\end{equation}
where
\begin{equation}
E^{int}(\beta,\mu_{1},\ldots,\mu_{M})=J\sum_{i=1}^{L}\sum_{m=1}^{M}<\vec{S}_{im}^{e}\cdot
\vec{S}_{i}^{f}>_{i}
\end{equation}
and
\begin{eqnarray}
<A>_{i}&=&TrA\rho_{i}\nonumber\\
\rho_{i}&=&\frac{e^{-\beta H_{i}}}{Tre^{-\beta H_{i}}}\nonumber\\
H_{i}&=&\sum_{m=1}^{M}\left[J\vec{S}_{im}^{e}\cdot
\vec{S}_{i}^{f}+(t-\mu_{m})\hat{N}_{im}^{e}\right].
\end{eqnarray}
We are thus left with a system of spins with only on-site interaction.

At this point several definitions of the ground-state can be introduced. In the
grand-canonical ensemble, taking $\beta\rightarrow\infty$ with
$(\mu_{1},\ldots,\mu_{M})$ fixed, we obtain only some specific values for the
electron density $n$. For example for a single band and $J>0$, we have
\begin{equation}
\begin{array}{lll}
n=0 & \hspace{1cm}{\rm for} & \hspace{1cm}\mu<t-\frac{3J}{4}\\
n=\frac{1}{3} & \hspace{1cm}{\rm for} & \hspace{1cm}\mu=t-\frac{3J}{4}\\
n=1 & \hspace{1cm}{\rm for} & \hspace{1cm}\mu\in]t-\frac{3J}{4},t+\frac{3J}{4}[\\
n=\frac{5}{3} & \hspace{1cm}{\rm for} & \hspace{1cm}\mu=t+\frac{3J}{4}\\
n=2 & \hspace{1cm}{\rm for} & \hspace{1cm}\mu>t+\frac{3J}{4}.\\
\end{array}
\end{equation}

In the canonical ensemble we can consider the ground-states either for
fixed $(N_{1},\ldots,N_{M})$, or for fixed $N=\sum_{m=1}^{M}N_{m}$, with respect
to the hamiltonian
\begin{eqnarray}
H&=&-2t'LM\theta(t')+t\sum_{m=1}^{M}N_{m}+E^{int}\nonumber\\
E^{int}&=&J\sum_{i=1}^{L}\sum_{m=1}^{M}\vec{S}_{im}^{e}\cdot\vec{S}_{i}^{f}.
\end{eqnarray}
Therefore we only need to compute the ground-state energy of the spin-exchange
interaction, which we shall now obtain explicitly for one and two bands.

For a single band, we have to diagonalize the on-site exchange
operator $\vec{S}_{i}^{e}\cdot\vec{S}^{f}_{i}$.
For a given site $i$, the size of the Hilbert space is 8. The four basis states 
with zero or two electrons on this site have energy zero. For the four states
with one electron we have 
a triplet with energy $J/4$ and a singlet with 
energy $-3J/4$. Using these results, we
recover immediately the ground-state energy obtained in \cite{gruber},
both for the antiferromagnetic $(J>0)$
and the ferromagnetic $(J<0)$ case, i.e. for the antiferromagnetic case:
\begin{equation}
E^{int}_{GS}=\left\{\begin{array}{ll}
-\frac{3J}{4}N & \hspace{1cm}N\leq L \\
-\frac{3J}{4}(2L-N) & \hspace{1cm}N\geq L,
\end{array}\right.
\label{antiferro}
\end{equation}
for the ferromagnetic case:
\begin{equation}
E^{int}_{GS}=\left\{\begin{array}{ll}
\frac{J}{4}N & \hspace{1cm}N\leq L \\
\frac{J}{4}(2L-N) & \hspace{1cm}N\geq L.
\end{array}\right.
\label{ferro}
\end{equation}

For the two bands case, we have to diagonalize the on-site operator 
$\vec{S}_{i}^{e}(1)\cdot\vec{S}_{i}^{f}+\vec{S}_{i}^{e}(2)\cdot\vec{S}_{i}^{f}$.
The size of the Hilbert space for the site $i$ is 32. The 8 states with either
zero or two electrons in each band at site $i$ have energy zero since the electronic spins
are zero. For the 16 states with one band occupied by zero or two electrons and
the other band occupied by one electron, we have again a triplet with energy
$J/4$ and a singlet with energy $-3J/4$, since the energy for the band with zero
or two electrons is zero.  
Finally, for the 8 states with exactly
one electron in each band, we have a quadruplet with energy  $J/2$, a doublet
with energy  $-J$ and a doublet with energy zero.
The ground-state energy $E^{int}_{GS}$ for the system with $N_{1}$ electrons in band
1 and $N_{2}$ electrons in band 2 can thus be computed explicitly and is given
by $E_{\alpha}$ for $(N_{1},N_{2})$ in the domain ${\cal D}_{\alpha}$,
$\alpha=1,\ldots,8$, represented in figure 1 below. 

\begin{figure}[htbp]
\begin{center}
\mbox{\psfig{figure=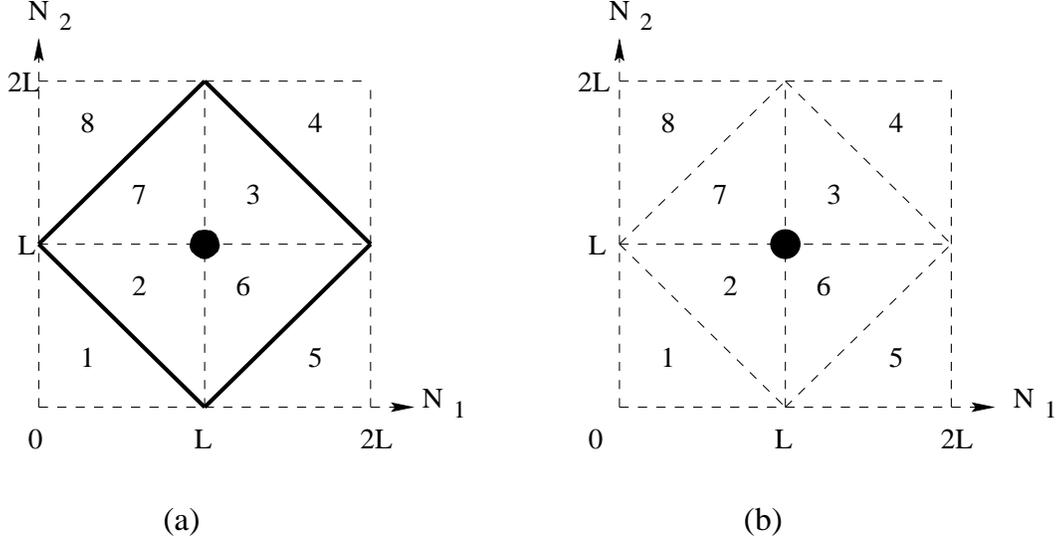,width=14cm}}
\end{center}
\caption{Ground-state energy for $J>0$ (a) and for $J<0$ (b). Solid lines
indicate filling numbers where phase transitions occur.}
\end{figure}

For the antiferromagnetic case $(J>0)$, the ground-state energies are given by
\begin{equation}
\begin{array}{lcl}
E_{1}=-\frac{3}{4}J(N_{1}+N_{2})&\hspace{1cm}&E_{5}=-\frac{3}{4}J(2L-N_{1}+N_{2})\\
E_{2}=-\frac{1}{4}J(2L+N_{1}+N_{2})&\hspace{1cm}&E_{6}=-\frac{1}{4}J(4L-N_{1}+N_{2})\\
E_{3}=-\frac{1}{4}J(6L-N_{1}-N_{2})&\hspace{1cm}&E_{7}=-\frac{1}{4}J(4L+N_{1}-N_{2})\\
E_{4}=-\frac{3}{4}J(4L-N_{1}-N_{2})&\hspace{1cm}&E_{8}=-\frac{3}{4}J(2L+N_{1}-N_{2})\\
\end{array}
\end{equation}
and for the ferromagnetic case $(J<0)$
\begin{equation}
\begin{array}{lcl}
E_{1}=E_{2}=\frac{1}{4}J(N_{1}+N_{2})&\hspace{1cm}&E_{5}=E_{6}=\frac{1}{4}J(2L-N_{1}+N_{2})\\
E_{3}=E_{4}=\frac{1}{4}J(4L-N_{1}-N_{2})&\hspace{1cm}&E_{7}=E_{8}=\frac{1}{4}J(2L+N_{1}-N_{2}).\\
\end{array}
\end{equation}
The differences between the two cases $J>0$ and $J<0$ can be seen as follows. In
the domain ${\cal D}_{1}$ where $N_{1}+N_{2}\leq L$, the lowest energy states
are obtained with zero or one electron at each site in any of the two bands,
which gives $E_{1}=-\frac{3}{4}JN$ if $J>0$ and $E_{1}=\frac{1}{4}N$ if $J<0$.
On the other hand in the domain ${\cal D}_{2}$ we have $(2L-N)$ sites with one
electron and $(N-L)$ sites which are doubly occupied with one electron in each
band and total electron spin 1. In this domain ${\cal D}_{2}$, if $J>0$ the
total electrons plus impurity spin is $1/2$, the three spins form a doublet, and
the energy is $E_{2}=-\frac{J}{4}(2L+N)$; in this case
$dE^{int}_{GS}/dN$ is discontinuous at $N=L$. However if $J<0$, the
total electrons plus impurity spin is $3/2$, the three spins  form a quadruplet,
and the energy is $E_{2}=\frac{1}{4}JN$; in this case there is no discontinuity
of $dE^{int}_{GS}/dN$
at $N=L$. It is interesting to note that even though the two electronic spins
are not directly coupled in the hamiltonian, they do interact indirectly via the
impurity spin. 
For a system with $N_{1}+N_{2}=N$ fixed, it is easily seen that the ground-state
is obtained for $(N_{1},N_{2})$ in the domains ${\cal D}_{\alpha}$, $\alpha=1,2,3,4.$

The exact expressions for the ground-state energies allow us to study the
conducting properties of the system at zero temperature.
For the single band case, following an idea due to Mattis \cite{mattis,lieb}, we define
$\mu^{+}=E_{GS}(N+1)-E_{GS}(N)$, $\mu^{-}=E_{GS}(N)-E_{GS}(N-1)$ and $\Delta\mu=\mu^{+}-\mu^{-}$. If
$\Delta\mu=0$, the system is conducting; if $\Delta\mu>0$, the system is
insulating due to the apparition of a gap $\Delta\mu$. For a single band, it
follows from (\ref{antiferro}) and (\ref{ferro}) that $\Delta\mu=0$ except for
$N=L$ where $\Delta\mu=3J/2$ if $J>0$ and $\Delta\mu=-J/2$ if $J<0$; therefore,
as shown in \cite{gruber},
the system is always conducting except at the point $N=L$ where
it is insulating.
For the two bands case, if we consider the ground-states with $N_{1}$ and
$N_{2}$ fixed, we have to introduce two parameters $\Delta\mu_{1}$ and
$\Delta\mu_{2}$ referring to the two bands,
\begin{eqnarray}
\Delta\mu_{1}=E_{GS}(N_{1}+1,N_{2})-2E_{GS}(N_{1},N_{2})+E_{GS}(N_{1}-1,N_{2})\nonumber\\
\Delta\mu_{2}=E_{GS}(N_{1},N_{2}+1)-2E_{GS}(N_{1},N_{2})+E_{GS}(N_{1},N_{2}-1).
\end{eqnarray}
The system is insulating if and only if a gap appears in the two bands,
i.e. if $\Delta\mu_{1}>0$ and $\Delta\mu_{2}>0$.

For the antiferromagnetic case, the system is insulating at the point $N_{1}=N_{2}=L$
where $\Delta\mu_{1}=\Delta\mu_{2}=\frac{J}{2}$ and on the four lines
$N_{1}+N_{2}=L$, $N_{1}-N_{2}=L$, $N_{2}-N_{1}=L$, $N_{1}+N_{2}=3L$ where 
$\Delta\mu_{1}=\Delta\mu_{2}=\frac{J}{2}$. We can thus conclude that a
metal-insulator transition occurs on the four heavy lines of figure 1(a) and at the point
$N_{1}=N_{2}=L$, as we turn on the antiferromagnetic interaction between
electrons and impurities.
For the ferromagnetic case, the system is insulating only at the point
$N_{1}=N_{2}=L$ where $\Delta\mu_{1}=\Delta\mu_{2}=-\frac{J}{2}$  and thus at
this point a metal-insulator transition occurs as the ferromagnetic interaction
is introduced.
Similarly, if we consider the ground-states with $N_{1}+N_{2}=N$ fixed, then for
$J>0$ the system is insulating if $N=L,2L,3L$, while for $J<0$ it is insulating
only for $N=2L$.

\section{Ground-state wavefunctions at $J=\infty$}

In this section, we consider a finite system $\Lambda$ with $L$ sites and an arbitrary
number $M$ of electronic bands. We want to obtain the ground-state wavefunctions
for the case $t<0$ and in the limit $J=\infty$.

If the total number of electrons is smaller than the number of sites, i.e.
\begin{equation} 
N=\sum_{m=1}^{M}N_{m}<L,
\end{equation}
then each electron will attempt to form a singlet of energy
$-3J/4$ with an impurity. Therefore in the limit $J=\infty$ we can reduce the Hilbert
space to the subspace where at each site there is either a singlet or an
unpaired impurity spin. A general basis for this subspace is given by the
following vectors
\begin{eqnarray}
\mid\alpha>&=&\mid X_{1},\ldots,X_{M};\sigma>\nonumber\\
&=&\prod_{m=1}^{M}\left[\prod_{x\in X_{m}}\frac{1}{\sqrt{2}}(c_{x\uparrow m}^{+}f_{x\downarrow}^{+} 
-c_{x\downarrow m}^{+}f_{x\uparrow}^{+})\right]\prod_{n=1}^{L-N}
f_{y_{n},\sigma_{y_{n}}}^{+}\mid 0>
\end{eqnarray}
where the $N_{m}$ singlets with electron in the band $m$ are located at the
positions $X_{m}=(x_{1m},\ldots,x_{N_{m}m})$, $X_{m}\cap X_{m'}=\emptyset$,
and the unpaired impurity spins $\sigma=(\sigma_{1},\ldots,\sigma_{L-N})$
are located at the positions $Y=\Lambda\setminus\cup_{m}X_{m}=(y_{1},\ldots,y_{L-N})$  
with $y_{1}<y_{2}<\ldots<y_{L-N}$.

With $P$ the projector onto this $J=\infty$ subspace, the projected hamiltonian 
has the form
\begin{equation}
PHP=PH_{0}P-\frac{3J}{4}N
\end{equation}
where $H_{0}$ is the kinetic term. Moreover, we can then forget the (infinite)
constant and consider only the term $PH_{0}P$. 
To find the ground-state wavefunctions, we identify the singlets as
spinless bosons and the unpaired impurities as spin $1/2$ fermions.
Let us then introduce the hamiltonian
\begin{equation}
h=-\frac{t}{2}\sum_{i\neq j=1}^{L}\sum_{\sigma=\uparrow,
\downarrow}\sum_{m=1}^{M}P_{G}F_{i\sigma}B_{im}^{+}
F_{j\sigma}^{+}B_{jm}P_{G}
\label{newham}
\end{equation}
where the $B$ fields are bosonic, and the $F$ fields are fermionic. The $B$
fields commute with the $F$ fields and $\sum_{i=1}^{L}B_{im}^{+}B_{im}=N_{m}$.
The Gutzwiller projector $P_{G}$ projects on the subspace where at each site
there is exactly one particle, either one fermion or one boson, i.e.
\begin{equation}
\sum_{m=1}^{M}B_{im}^{+}B_{im}+\sum_{\sigma}F_{i\sigma}^{+}F_{i\sigma}=1.
\end{equation}
For this system, the basis vectors can be taken as follows
\begin{equation}
\mid\bar {\alpha}>=\prod_{m=1}^{M}\left[\prod_{x\in X_{m}}
B_{xm}^{+}\right]\prod_{n=1}^{L-N}F_{y_{n},\sigma_{y_{n}}}^{+}
\mid 0>.
\end{equation} 
The two systems defined respectively by the hamiltonians $PH_{0}P$ and $h$ are
isomorphic, since there is a one to one correspondence
$\mid\alpha>\rightarrow\mid\bar{\alpha}>$ between the basis vectors and the
matrix elements of the two hamiltonians are the same
\begin{equation}
<\beta\mid PH_{0}P\mid\alpha>=<\bar{\beta}\mid h \mid\bar{\alpha}>.
\end{equation}
 
As for the Hubbard model \cite{verges,brandt}, we try to write the hamiltonian as a
positive definite form to determine the ground-state energy. Using the relations
\begin{eqnarray}
P_{G}F_{i\sigma}^{+}B_{im}F_{j\sigma}B_{jm}^{+}P_{G}&=&
-F_{j\sigma}B_{jm}^{+}P_{G}F_{i\sigma}^{+}B_{im}\;\;{\rm if}\;i\neq j\nonumber\\
P_{G}B_{im}^{+}B_{im}P_{G}&=&F_{i\sigma}B_{im}^{+}P_{G}F_{i\sigma}^{+}B_{im}
\end{eqnarray}
we can rewrite the hamiltonian $h$ in the following way:
\begin{equation}
h=-\frac{t}{2}\sum_{i,j=1}^{L}\sum_{\sigma=\uparrow,\downarrow}\sum_{m=1}^{M}
F_{i\sigma}B_{im}^{+}P_{G}
F_{j\sigma}^{+}B_{jm}+tP_{G}\sum_{m=1}^{M}\hat{N}_{m}P_{G}.
\end{equation}
The first part of $h$ is positive definite since $t<0$, and the second part
is a constant. 

If $N=\sum_{m}N_{m}\leq L-2$, the ground-state wavefunctions are labelled by the
positions $\bar{Y}=(y_{1},\ldots,y_{Q})$ and the spins
$\bar{\sigma}=(\sigma_{1},\ldots,\sigma_{Q})$ of $L-N-2=Q$ fermions. These
wavefunctions are given by
\begin{equation}
\mid\Psi_{GS}(\bar{Y},\bar{\sigma})>=\sum_{z_{1},z_{2}}
\sum_{X_{1},\ldots,X_{M}\atop
\cup_{m}X_{m}\cup\{z_{1},z_{2}\}=\Lambda\setminus\bar{Y}}
F_{z_{1}\uparrow}^{+}
F_{z_{2}\downarrow}^{+}\prod_{m=1}^{M}\left[\prod_{x_{m}\in
X_{m}}B_{x_{m}m}^{+}\right]\prod_{n=1}^{Q}
F_{y_{n}\sigma_{y_{n}}}^{+}\mid 0>
\label{wf}
\end{equation}
where in Eq.(\ref{wf}) we sum over the positions $z_{1},z_{2}$ of the two last
fermions with opposite spin, and over the positions $X_{1},\ldots,X_{M}$ of the
bosons in each band:
$$X_{m}\cap X_{m'}=\emptyset\;\;{\rm if}\;m\neq m'$$
$$X_{m}\cap \{z_{1},z_{2}\}=\emptyset$$
$$z_{1}\neq z_{2}.$$
To prove that (\ref{wf}) is indeed a ground-state, one shows that it is
annihilated by the first part of $h$ and therefore (\ref{wf}) is
a ground-state wavefunction with energy tN.

If $N=\sum_{m}N_{m}=L-1$, the ground-state wavefunctions have only one delocalized fermion
and are parametrized by the spin $\sigma$ of this
fermion. They are given by
\begin{equation}
\mid\Psi_{GS}^{\sigma}>=\sum_{z}\sum_{X_{1},\ldots,X_{M}\atop
\cup_{m}X_{m}\cup\{z\}=\Lambda}F_{z\uparrow}^{+}
\prod_{m=1}^{M}\left[\prod_{x_{m}\in X_{m}}B_{x_{m}m}^{+}\right]\mid 0>
\end{equation}
with 
$$X_{m}\cap X_{m'}=\emptyset\;\;{\rm if}\;m\neq m'$$
$$X_{m}\cap \{z\}=\emptyset$$
and the energy is $\frac{t}{2}(L-1)$.

As for the single band case \cite{gruber},
if we impose periodic boundary conditions, we can write the ground-state
wavefunctions in a Jastrow product form.
Let us consider a state with $N_{+}$
fermions with up spin localized at $Y^{+}=(y^{+}_{1},\ldots,y^{+}_{N_{+}})$, 
$N_{-}$ fermions with down spin localized at $Y^{-}=(y^{-}_{1},\ldots,y^{-}_{N_{-}})$
and $Q=L-N_{+}-N_{-}$ holes localized at $X=(x_{1},\ldots,x_{Q})$.
The ground-state wavefunctions are given by
\begin{eqnarray}
\mid\Psi_{GS}>=\sum_{X,Y^{-}}\Psi(X,Y^{-})
&&\prod_{m=1}^{M}\left[\prod_{j=\sum_{k=1}^{m-1}N_{m}}
^{\sum_{k=1}^{m}N_{m}}
B_{x_{j}m}^{+}F_{x_{j}\uparrow}\right]\cdot\nonumber\\
& &\cdot\prod_{j=1}^{B}F_{Y^{-}_{j}\downarrow}^{+}F_{Y^{-}_{j}\uparrow}\prod_{n=1}^{L}
F_{n\uparrow}^{+}\mid 0>
\end{eqnarray}
with amplitude
\begin{equation}
\Psi(X,Y^{-})=e^{\frac{2\Pi i}{L}(m_{h}\sum_{i}x_{i}+m_{s}\sum_{j}y^{-}_{j})}
\prod_{i<j}d(x_{i}-x_{j})\prod_{i,j}d(y^{-}_{i}-x_{j})\prod_{i<j}d^{2}(y^{-}_{i}-y^{-}_{j})
\end{equation}
where the function $d(n)=\sin(n\pi/L)$ and the quantum numbers
$m_{h}$ and $m_{s}$ are integers or half-integers which make sure of the 
periodic boundary conditions and satisfy the following inequalities
\begin{equation}
\frac{1}{2}(N_{-}+Q+1)\leq m_{h} \leq L-\frac{1}{2}(N_{-}+Q+1)
\end{equation}
\begin{equation}
\frac{1}{2}(N_{+}+Q+1)\leq m_{h}-m_{s}+\frac{L}{2} \leq
L-\frac{1}{2}(N_{+}+Q+1).
\end{equation}

To see that such states are ground-states, one has to verify
that they are eigenstates of the hamiltonian (\ref{newham}) with eigenvalues
$tN$. To establish this result, one has first to apply the up-spin part
of this hamiltonian and see that $\mid\Psi_{GS}>$ is an eigenvector with eigenvalue 
$\frac{t}{2}N$. To apply the down-spin part of the
hamiltonian, it is more convenient to write the ground-state in terms of
holes and up-spins \cite{wang}. In these terms, the amplitude $\Psi(X,Y^{+})$
has the same form as $\Psi(X,Y^{-})$ except that $m_{h}$ is replaced by  
$m_{h}-m_{s}+L/2$ and $m_{s}$ by $L-m_{s}$. In this manner, it gives an energy 
$\frac{t}{2}N$ and the total energy is $tN$.

\section{Summary}
  
The ground-state energy of our Kondo-lattice model was obtained explicitly in
the thermodynamic limit for one and two electronic bands. From this solution
the insulating or conducting properties of the system were established as a
function of the number of electrons in each bands. For the ferromagnetic system
with two electronic bands a metal-insulator phase transition appears at
$N_{1}=N_{2}=L$, as the interaction between the electrons and the impurities is
switched on at zero temperature. For the antiferromagnetic system, new
metal-insulator phase transitions also appear for other values of $N_{1}$ and $N_{2}$.

For the finite size system, we have shown that the ground-state wavefunctions
can be written in the well-known Jastrow product form.
It remains unclear how to define the effective masses for electrons 
hopping with an unconstrained hopping amplitude and whether the impurity spins 
induce heavy masses for the conduction electrons. 

This work was supported in part by the Swiss National Science Foundation.


\end{document}